\documentclass[twocolumn,aps,prb,showpacs,letterpaper]{revtex4}
\usepackage{amsmath}
\usepackage{graphicx}
\usepackage{hyperref}

\def\Re{\operatorname{Re}}
\def\Im{\operatorname{Im}}
\def\Tr{\operatorname{Tr}}
\def\k{\mathbf{k}}
\def\p{\mathbf{p}}
\def\q{\mathbf{q}}
\def\v{\mathbf{v}}
\def\j{\mathbf{j}}
\def\O{\mathcal{O}}
\def\eps{\varepsilon}

\begin{document}

\title{Supercurrent in Nodal Superconductors}

\author{Igor Khavkine$^1$, Hae-Young Kee$^1$}
\author{K. Maki$^2$}
\affiliation{${}^1$Department of Physics,
	University of Toronto, Toronto, Ontario M5S 1A7, Canada\\
	${}^2$Department of Physics and Astronomy, University of Southern California,
	Los Angeles, CA 90089-0484}
\date{\today}

\begin{abstract}
In recent years, a number of nodal superconductors have been identified;
$d$-wave superconductors in high $T_c$ cuprates, CeCoIn$_5$, and
$\kappa$-(ET)$_2$Cu(NCS)$_2$, 2D $f$-wave superconductor in
Sr$_2$RuO$_4$ and hybrid $s{+}g$-wave superconductor in YNi$_2$B$_2$C.
In this work we conduct a theoretical study of nodal superconductors in
the presence of supercurrent. For simplicity, we limit ourselves to
$d$-wave  and 2D $f$-wave superconductors.  We compute the quasiparticle
density of states and the temperature dependence of the depairing
critical current in nodal superconductors, both of which are accessible
experimentally.
\end{abstract}


\maketitle

\section{Introduction}
Since the discovery of the organic (TMTSF)$_2$PF$_6$ and heavy fermion
CeCu$_2$Si$_2$ superconductors in 1979,
the high $T_c$
cuprate superconductors LaBaCuO$_4$ in 1986 and Sr$_2$RuO$_4$ and
YNi$_2$B$_2$C in 1994, the new class of nodal superconductors took
center stage.\cite{won1} However, until recently, except for $d$-wave
symmetry in high $T_c$ cuprate superconductors,\cite{tsuei,damascelli}
the gap symmetry has not been explored directly for other nodal
superconductors.

In the last few years, it has been recognized that the Doppler shift in
the quasiparticle spectrum in the vortex state\cite{volovik, kubert}
provides extremely sensitive means of studying the nodal lines and
points in the energy gap $\Delta(\k)$.\cite{vekhter,won2,dahm,won3} More
recently the above analysis has been extended to the hybrid $s{+}g$-wave
superconductors.\cite{maki1,maki2, maki3}
Indeed, through angular dependent magnetothermal conductivity
measurements on high quality single crystals at low temperatures ($T <
1\mathrm{K}$), Izawa \emph{et al.}\ have identified 2D $f$-wave
superconductivity in Sr$_2$RuO$_4$,\cite{izawa1} $d$-wave
superconductivity in CeCoIn$_5$\cite{izawa2} and
$\kappa$-(ET)$_2$Cu(NCS)$_2$,\cite{izawa3} as well as $s{+}g$-wave
superconductivity in YNi$_2$B$_2$C.\cite{izawa4}

However, as seen in Ref.~\onlinecite{mackenzie}, the nature of
superconductivity in Sr$_2$RuO$_4$ is still controversial.  Clearly both
the specific heat data and the magnetic penetration depth data indicate
the presence of line nodes in
Sr$_2$RuO$_4$.\cite{dahm,nishizaki,bonalde} Further, both the angular
dependent thermal conductivity data\cite{izawa1} and the ultrasonic
attenuation data\cite{lupien} exclude models with vertical line
nodes.\cite{miyake,graf}  Finally the thermal conductivity data by Izawa
\emph{et al.}\cite{izawa1}\ and Suzuki \emph{et al.}\cite{suzuki}\ are
incompatible with the two gap model by Zhitomirsky and
Rice.\cite{zhitomirsky,annett} Therefore  it is important to understand
whether $f$-wave superconductivity is realized in Sr$_2$RuO$_4$.  More
recent angular dependent magnetospecific heat data for $T>0.15K$ by
Deguchi \emph{et al.}\cite{deguchi}\ do not change this situation,
though some of the horizontal nodes may be converted into a set of point
minigaps.\cite{won-jang1}

In this paper, we concentrate on two superconducting
order parameters\cite{won1}
\begin{align}
	\Delta(\k) &= \Delta \cos(2\phi) & \text{$d$-wave,} \\
	\hat{\Delta}(\k) &=	\hat{d} \Delta \exp(\pm i \phi) \cos(\chi)
		& \text{$f$-wave,}
\end{align}
where $\phi$ is the in-plane angle measured from the $a$-axis ($\phi
=\tan^{-1} (k_y/k_x)$), and $\chi = c k_z$ with $c$, being the $c$-axis
lattice constant.
We study the quasiparticle density of states (DOS) in the presence of
superconductivity (SC) and external current, which is accessible through
tunneling spectroscopy. We also examine the dependence of critical
current in the superconductor as a function of temperature.

Making use of the mean field quasiparticle Green function we can express
the quasiparticle density of states in the presence of uniform
supercurrent in terms of simple integrals which are evaluated
numerically. The superconducting order parameter is also determined
within the mean field theory. Then, as in the $s$-wave
superconductor,\cite{maki4} the order parameter $\Delta(T)$ is modified
in the presence of supercurrent. We see that the supercurrent first
increases linearly with the pair momentum $\q_s$, reaches a maximum
value, and drops as $\q_s$ is further increased. We call this maximum
value the \emph{depairing critical current} in contrast to the usual
critical current associated with depinning of vortices and vortex
lattices.
The advantage of studying the supercurrent, compared to other transport
properties, lies in its sensitivity to the condensate itself.

\section{Quasiparticle Density of States}
In the presence of superflow, the quasiparticle Green function in the
Nambu-Gor'kov formalism is given by\cite{maki4,kee,won4}
\begin{equation}
	G^{-1}(i\omega_n, \k) = i\omega_n - \v_F\cdot\q_s
		+ \xi_k \rho_3 + \Delta(\k) \rho_1 \sigma_1,
\end{equation}
where $\sigma_i$ and $\rho_i$ are Pauli matrices acting on the spin and
Nambu indices respectively.
In the quasi-2D system we are considering\cite{won-maki1}
\begin{equation}
	\xi_k = \frac{1}{2m} (k_x^2+k_y^2) - 2 t \cos\chi -\mu
\end{equation}
and $\q_s$ is the pair momentum.

The quasiparticle density of states in the presence of supercurrent is
given by
\begin{equation}
	N(E) = -\frac{1}{\pi} \sum_\k \Im \Tr[G(E,\k)],
\end{equation}
which can be reduced to
\begin{equation}
	g(E) \equiv N(E)/N_0 = \Re \left\langle
		\frac{|E-\v_F\cdot\q_s|}{\sqrt{(E-\v_F\cdot\q_s)^2-\Delta^2 f^2}}
	\right\rangle,
\end{equation}
where $N_0 = 2m/\pi$, the function $f$ is $\cos(2\phi)$ for the
$d$-wave and $\cos(\chi)$ for the $f$-wave cases, and the averaging is
done over angles $\langle {\cdots}\rangle \equiv \frac{1}{(2\pi)^2}\int
d\phi \int d\chi\, ({\cdots})$.  For simplicity we shall consider the
following four cases.


\subsection{$f$-wave SC with $J \parallel a$
	and $d$-wave SC with $J \parallel c$\label{fja}}
The quasiparticle density of states (DOS) is given by
\begin{equation}\label{fja-dos-eq}
 g(E,s) = \frac{1}{\pi}\int_{0}^{\pi}d\phi\, g_d(E,s,\phi),
\end{equation}
where $s = v_F q_s$ for in-plane current or $s = 2t q_s\equiv
v_{Fc}q_s$ for $c$-axis current, and
\begin{equation}
	g_d(E,s,\phi) = \frac{2}{\pi} \Re\left\{ K\left(
		\frac{\Delta}{|E-s\cos\phi|} \right) \right\}.
\end{equation}
The quasiparticle density of states in the absence of supercurrent
for both $f$-wave and $d$-wave cases is given by $g_d(E,0,0)$.\cite{won4}
Here $K(k)$ is the Complete Elliptic Integral of the First Kind as a
function of the modulus\cite{JE} $k$ and
$\Re\{K(k)\} = k^{-1} K(k^{-1})$ for $k>1$.

\begin{figure}
	\includegraphics[width=8cm]{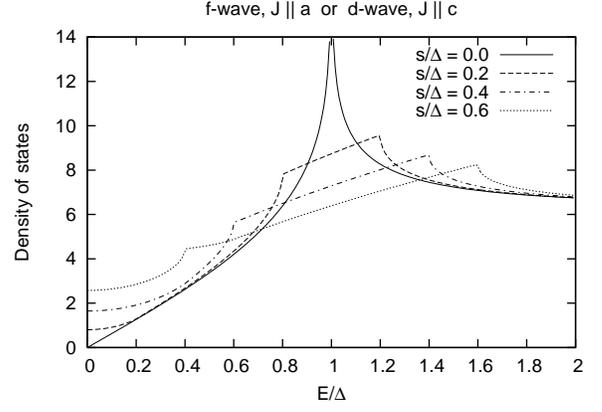}
	\caption{Quasiparticle DOS $g(E,s)$ given in
		Eq.~\eqref{fja-dos-eq}.\label{fja-dos}}
\end{figure}

The quasiparticle DOS is shown in Fig.~\ref{fja-dos}.  In the absence of
current, say for the $d$-wave case, the quasiparticle spectrum has a
line of degenerate saddle points along $\chi$ which give rise to a
logarithmic peak in the DOS. At finite current, the Doppler shift
$s\sin\chi$ breaks this degeneracy, leaving only two discrete saddle
points. Thus the logarithmic peak at $|E|=\Delta$ is split into two
cusps at $|E|=\Delta\pm s$.


\subsection{$f$-wave SC with $J \parallel c$\label{fjc}}
\begin{figure}
	\includegraphics[width=8cm]{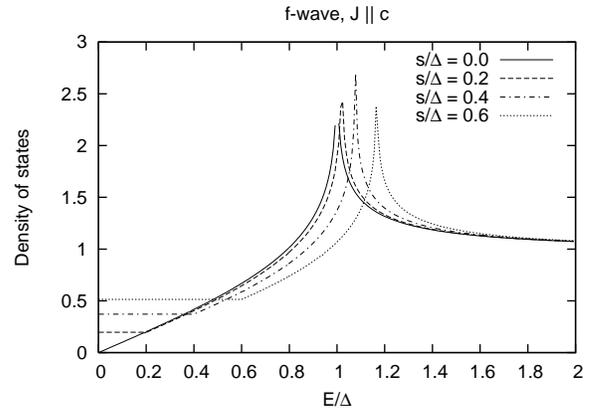}
	\caption{Quasiparticle DOS $g(E,s)$ given in
		Eq.~\eqref{fjc-dos-eq}.\label{fjc-dos}}
\end{figure}

The integral reduces to
\begin{equation}\label{fjc-dos-eq}
	g(E,s) = \frac{1}{\pi}\Re\int_{-\pi/2}^{\pi/2} d\chi
		\frac{|E-s\sin\chi|}{\sqrt{(E-s\sin\chi)^2 - \Delta^2\cos^2\chi}}.
\end{equation}
In the limit $E \rightarrow 0$, $g(0,s) =\frac{s}{\sqrt{s^2+\Delta^2}}$.
The quasiparticle DOS is shown in Fig.~\ref{fjc-dos}.  The Doppler shift
$s\sin\chi$ does not lift the degeneracy of the line of saddle points
in the quasiparticle spectrum along $\phi$, but does shift them. Thus
the logarithmic peak at $|E|=\Delta$ is shifted to
$|E|=\sqrt{\Delta^2+s^2}$. Also, the DOS flattens out for $|E|<s$ as
more low energy states are added near the gap nodes.

\subsection{$d$-wave SC with $J \parallel a$\label{dja}}
The DOS becomes 
\begin{equation}\label{dja-dos-eq}
	g(E,s) = \frac{1}{\pi} \Re \int_{0}^{\pi}d\phi
		\frac{|E-s\cos\phi|}{\sqrt{(E-s\cos\phi)^2-\Delta^2\cos^2(2\phi)}}.
\end{equation}
When $E=0$, we find
\begin{equation}\label{d_para_a}
	g(0,s) = \frac{s}{\pi\Delta} \Re\left\{ \frac{1}{\sqrt{a}}
		K\left[\sqrt{\frac{b}{a}}\right] \right\},
\end{equation}
where $b=s\sqrt{s^2+8\Delta^2}/(4\Delta^2)$ and $a =
(1+b)/2-s^2/(8\Delta^2)$.

\begin{figure}
	\includegraphics[width=8cm]{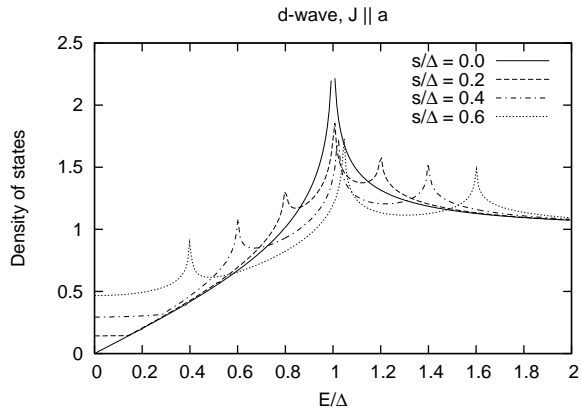}
	\caption{Quasiparticle DOS $g(E,s)$ given in
		Eq.~\eqref{dja-dos-eq}.\label{dja-dos}}
\end{figure}

The quasiparticle DOS is shown in Fig.~\ref{dja-dos}. Just as in the
previous section, the logarithmic peak at $|E|=\Delta$ is shifted to
$|E|=\sqrt{\Delta^2+s^2/4}$, while secondary peaks appear at
$|E|=\Delta\pm s$ due to the interference of the gap $\Delta\cos(2\phi)$
and Doppler shift $s\cos\phi$ modulations.

\subsection{$d$-wave SC with $J \parallel [110]$\label{dj110}}
The DOS is given by
\begin{equation}\label{dj110-dos-eq}
	g(E,s) = \frac{1}{\pi} \Re \int_{0}^{\pi} d\phi
		\frac{|E-s\cos\phi|}{\sqrt{(E-s\cos\phi)^2-\Delta^2\sin^2(2\phi)}}.
\end{equation}
For $E=0$, we find
\begin{equation}
	g(0,s) = \frac{2}{\pi} \Re\left\{ K \left(\frac{2\Delta}{s}\right)
	\right\}.
\end{equation}

\begin{figure}
	\includegraphics[width=8cm]{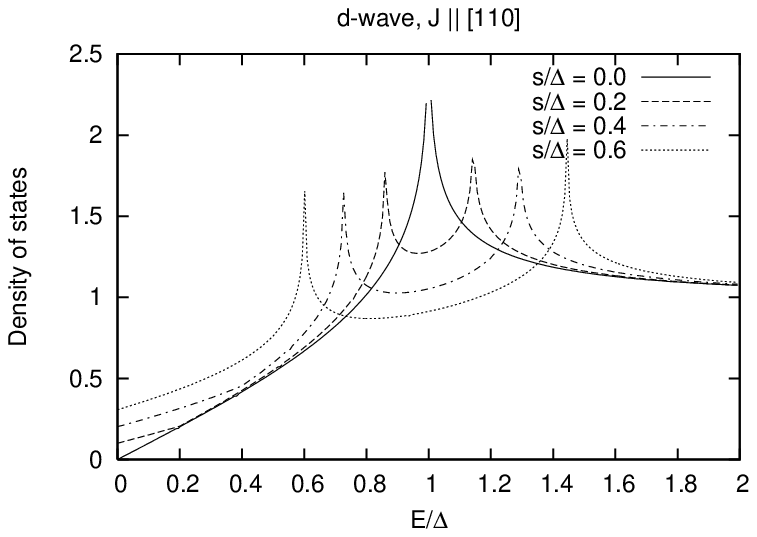}
	\caption{Quasiparticle DOS $g(E,s)$ given in
		Eq.~\eqref{dj110-dos-eq}.\label{dj110-dos}}
\end{figure}

The quasiparticle DOS is shown in Fig.~\ref{dj110-dos}.
As in the previous section, logarithmic singularities persist, but the
main peak at $|E|=\Delta$ is split into two smaller ones approximately
at $|E|=\Delta \pm s/\sqrt{2}+\O((s/\Delta)^2)$ for small $s/\Delta$.

\section{Gap equation and Critical current}
Within BCS theory, the gap equation is given by 
\begin{equation}
	\Delta(\k) = T\sum_{i\omega_n}\sum_\p V_{\k\p}
		\Tr[\rho_1\sigma_1 G(i\omega_n,\p)],
\end{equation}
where $\Delta(\k) \equiv \Delta f(\k)$ and $V_{\k\p} \equiv V f(\k)
f(\p)$. This equation can be reduced to
\begin{equation}\label{gapeq}
	-\frac{1}{2}\ln\left(\frac{\Delta}{\Delta_0}\right)
	= \Re \left\langle f^2 \cosh^{-1}\left(\frac{sz}{\Delta f}
	\right) \right\rangle
\end{equation}
at $T=0$, where $z=\cos(\phi-\phi_0)$ or $z=\sin\chi$ for the in-plane
or $c$-axis current respectively, with $\phi_0$ being the direction of
the in-plane current. $\Delta_0$ is the order parameter at $T=0$ and
$\q_s=0$. At zero temperature, the $f$-wave case was previously investigated in
Ref.~\onlinecite{kee}, while the $s$ and $d$-wave cases were considered
in Ref.~\onlinecite{zhang}.

\begin{figure}
	\includegraphics[width=8cm]{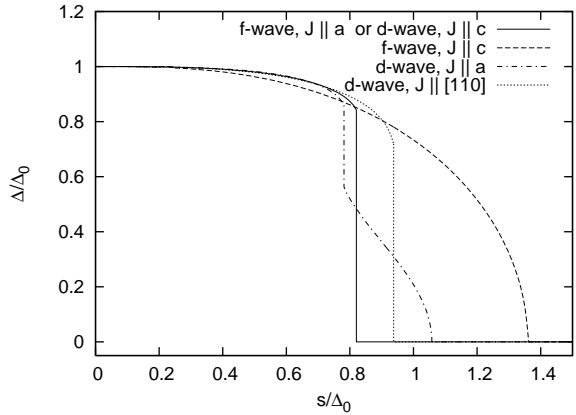}
	\caption{Zero temperature order parameter as function of $s$, as defined by
		Eq.~\eqref{gapeq}.\label{D_vs_y}}
\end{figure}

\begin{figure}
	\includegraphics[width=8cm]{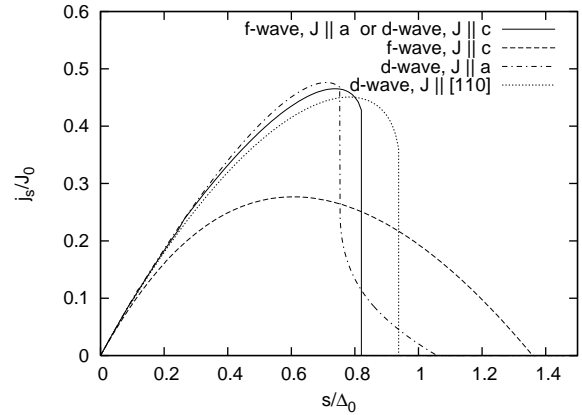}
	\caption{Zero temperature supercurrent as a function of $s$, given in
		Eq.~\eqref{jtot-T0}. Note that the $J_0$ differs by a factor of
		$v_{Fc}/v_F$ between the in-plane and $c$-axis cases.\label{j0_vs_y}}
\end{figure}

The dependence of the order parameter on $s$ is shown in
Fig.~\ref{D_vs_y}.
Unlike in the 3D $s$-wave superconductor, both the order parameter
$\Delta$ and the supercurrent $j_s$ jump (see Fig.~\ref{j0_vs_y}) at
$s/\Delta_0\sim 0.8$, except for the case of an $f$-wave
superconductor with $c$-axis current.
The values of $\Delta$ obtained via the gap equation and $j_s$ are shown
for $s$ increased from zero. In all cases the jump disappears at a finite
temperature and the transition to normal state becomes continuous. These
temperatures are $0.06\Delta_0$ and $0.25\Delta_0$ for the $d$-wave
superconductor with current along the $a$ and $[110]$ axes respectively.
The transition is continuous for all $T>0$ in the $f$-wave case with
in-plane current.

This behavior is strongly reminiscent of the case of Pauli paramagnetism in
both the $s$-wave and $d$-wave superconductors.\cite{maki4,won-jang2} These
jumps induced by the Pauli term signal the presence of more stable
inhomogeneous superconductivity, the Fulde-Ferrell-Larkin-Ovchinnikov (FFLO)
state.\cite{fulde,larkin} Very recently, presence of the FFLO state has been
reported from thermodynamic,\cite{bianchi} ultrasonic,\cite{watanabe} and
temperature dependent upper critical field measurements\cite{won-maki2} in
$d$-wave superconducting CeCoIn$_5$.\cite{izawa2}

Therefore, we speculate the appearance of inhomogeneous
superconductivity in the vicinity of $s/\Delta_0\sim 0.8$
similar to the FFLO state, but generated by uniform supercurrent, in
all the cases we have considered except for an $f$-wave superconductor with
$c$-axis current. A more detailed analysis of the free energy and
possible FFLO state will be reported in the near future.

The finite temperature gap equation reduces to
\begin{multline}\label{gapeq-T}
	-\frac{1}{2}\ln\left(\frac{\Delta}{\Delta_0}\right)
	= \Re \left\langle f^2 \cosh^{-1}\left(\frac{sz}{\Delta}
		\right) \right\rangle \\
		{}+ \left\langle f^2 \int_{\Delta|f|}^{\infty}dE\,
			\frac{h(E+sz)+h(E-sz)}{\sqrt{E^2-\Delta^2 f^2}} \right\rangle,
\end{multline}
where $h(\eps)=n_F(\eps)-n_F(\eps,T=0)$, with $n_F(\eps)$ being the
Fermi distribution.




Finally, we study the behavior of the supercurrent, which
can be evaluated as follows
\begin{equation}
	\j_s = \frac{ne\q_s}{m}
		+ T\sum_{i\omega_n}\sum_\k e\v_\k
			\Tr[G(i\omega_n,\k)],
\end{equation}
and reduces to
\begin{equation}\label{jtot-T0}
	j_s = J_0 s \left(1-2\Re \left\langle
		z^2\sqrt{1-\frac{\Delta^2 f^2}{(sz)^2}} \right\rangle \right)
\end{equation}
at $T=0$, where $J_0$ is a constant differing by a factor of
$v_{Fc}/v_F$ between the in-plane and $c$-axis cases. Solving
Eq.~\eqref{gapeq-T} for given $s$, and substituting the result into
Eq.~\eqref{jtot-T0}, we can also evaluate $j_s$ for given $s$.  The
behavior of the supercurrent as a function of $s$ is shown in
Fig.~\ref{j0_vs_y}. For the $f$-wave case, an analytic expression for
the zero temperature supercurrent and the depairing current
$j_{\text{max}}$ (at which $dj_s/ds = 0$) have been evaluated in
Ref.~\onlinecite{kee}. Also, the current curves for the $s$ and $d$-wave
cases were numerically evaluated in Ref.~\onlinecite{zhang}. As
expected, for large $s$ the equilibrium value of the current is zero
since superconductivity is destroyed. For small $s$, the current rises
linearly, it then reaches a maximum, and finally decreases to zero in
the same way as the order parameter.
%

\begin{figure}
	\includegraphics[width=8cm]{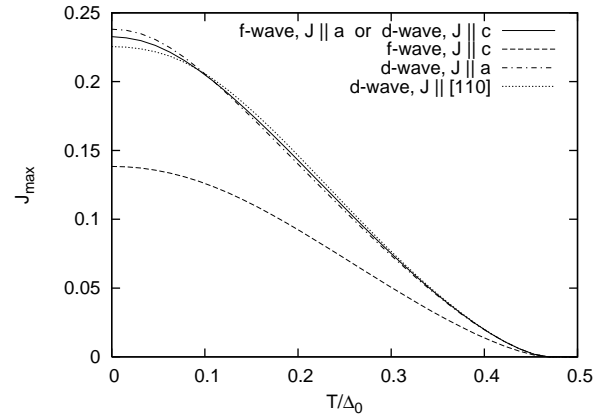}
	\caption{Depairing critical current as a function of temperature,
		defined by the maximum value of $j_s$ (Eq.~\eqref{jtot-T}) for given
		$T$.\label{j_max}}
\end{figure}

The finite temperature expression for the current becomes
\begin{multline}\label{jtot-T}
	j_s = J_0 s\left(1-2\Re \left\langle 
		z^2\sqrt{1-\frac{\Delta^2 f^2}{sz^2}} \right\rangle\right. \\
		\left. {}+ 2\left\langle z^2\int_{\Delta|f|}^\infty
			\frac{(h(E+sz)-h(E-sz))\,E\,dE}{\sqrt{E^2-\Delta^2 f^2}} \right\rangle
	\right).
\end{multline}
The dependence of the depairing current $j_{\text{max}}$ on temperature is
illustrated in Fig.~\ref{j_max}. The behavior of the critical current
as a function of temperature is qualitatively similar for all of the
considered cases except for the $f$-wave SC with current along the
$c$-axis. This case is set apart in the same way as when the behavior of
the order parameter was considered.


\section{Conclusion and Summary}
We consider the supercurrent as a simple pair breaking perturbation
which adds a Doppler shift to the quasiparticle spectrum.  While this
study is based on the assumption of uniform supercurrent throughout a
sample, the experimental realization of uniform current is far from
trivial.\cite{gray,hilgenkamp}
In bulk crystals, the supercurrent will be localized near the
sample surface due to the Meissner effect.  Therefore, it is important
to have high quality thin films or whiskers of relevant compounds.  Very
recently, whiskers of Bi2212 have been used to measure the cross
junction Josephson tunneling,\cite{takano,maki-haas} which implies that this
experiment, while nontrivial, should be possible in the future.

As long as the condition of uniform current is met, the quasiparticle
density of states obtained here will be readily accessible by scanning
tunneling microscopy and other tunneling techniques.  The temperature
dependence of the critical depairing current may also be measured.
These measurements will provide useful information on the pairing
symmetry of nodal superconductors.

As we have seen, the DOS is very sensitive to the symmetry of the gap
and the direction of supercurrent. In principle, the DOS is sensitive to
the location and type of critical points in the quasiparticle
dispersion.  These critical points are easily perturbed or displaced in
different ways by adding to the dispersion an appropriate term linear in
electron velocity---the Doppler shift. Thus, with the addition of a
uniform current, the change in the singularities of the DOS can reveal a
lot about the symmetry of the unperturbed quasiparticle dispersion.

Also, in many cases, the spatially homogeneous configuration as
considered here becomes unstable for $s/\Delta_0\sim 0.8$.
This opens up the possibility of a current induced
Fulde-Ferrell-Larkin-Ovchinnikov (FFLO) state. This possibility will be
further explored.

In summary, we have evaluated the effect of the presence of finite
supercurrent on the quasiparticle DOS of nodal superconductors.  The
quasiparticle DOS can be probed by tunneling spectroscopy and is
sensitive enough to the order parameter and supercurrent direction to
help differentiate between superconductors of different gap symmetries.
In addition, we showed the dependence of the supercurrent on the
external current and the dependence of the critical depairing current on
temperature.


\acknowledgments
This work was supported by NSERC of Canada (IK, HYK), Canada Research
Chair (HYK), Canadian Institute for Advanced Research (HYK), and Alfred P.\
Sloan Research Fellowship (HYK).

\end{document}